# Fault Localization in a Software Project using Back-Tracking Principles of Matrix Dependency

Vishal Anand[#1], Ramani S[*2]

[#]Student, SCSE, VIT University
[*]Assistant Professor, SCSE, VIT University
Vellore, Tamil Nadu, India

**ABSTRACT**

Fault identification and testing has always been the most specific concern in the field of software development. To identify and testify the bug we should be aware of the source of the failure or any unwanted issue. In this paper, we are trying to extract the location of failure and trying to cope up with the bug. Using directed graph, we tried to obtain the dependency of multiple activities in live environment to trace the origin of fault. Software development comes up with series of activities and we tried to show the dependency of multiple activities on each other. Critical activities are considered as they cause abnormal functioning of the whole system. The paper discuss about the priorities of activities of dependency of software failure on the critical activities. Matrix representation of activities as part of the software is chosen to determine root of the failure using concept of dependency. It can vary with the topography of network and software environment. When faults occur, the possible symptoms will be reflected in the dependency matrix with high probability in fault itself. Thus, independent faults are located in the main diagonal of dependency matrix.

**KEYWORDS**

*Software, Dependency, Matrix, Bug, Critical Activity, Modules*

## 1. INTRODUCTION

Developers unite to create software. A software project is initially handed over to a project manager. He then distributes the work to be done to developers. There are n number of attributes associated with a software project like- time of completion, cost of the project and the developers in that team. A software project during its development phase will comprise of several phases like- designing, coding, testing, debugging and maintenance [4]. This group of activities together is known as software development life cycle (SDLC).

### 1.1 Modular Structure of Project

Software projects are divided into modules to facilitate easy operation [10]. With the formation of modules, different teams are formed to perform different operations and then eventually integrate the work done by each team to form the desired software or to perform particular operation [6]. These fragmented units of work done by the teams are referred as modules and each module consists of multiple activities. Modularity in software environment helps to keep track of multiple activities going on to complete the project [7].

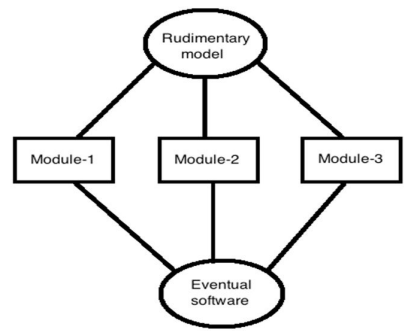

Fig. 1 Modular Approach

As explained modules consist of multiple activities that can be classified into three parts- critical activities, safe/non-critical activities, dummy activities. Critical activities are those, which are crucial to the working of the software and without which the functioning of the software would be at stake. Critical activities are indispensible part of software whereas; safe activities are those, which may be necessary but not mandatory at each and every step [8]. Even during any malfunction of safe activities, the software will perform although not impeccably. Critical activities decide the way of operation of the system as whole and failure of this result in malfunctioning of the entire system. On the other hand, we have non-critical activities that though consume time and resources but do not put any specific impact on system's overall architecture. Contrary to other activities, dummy activities does not consume any time and resource and mostly used in software project to maintain logical dependency and unique relationship by project managers [9].

## 2. METHODOLOGY

### 2.1 Critical Path Analysis

We are more concerned about critical activities as they form the significant part of software project. Sequence of critical activities in organised order forms Critical Path. Critical path activities are the project tasks that must start and finish on time to safeguard that the project ends on schedule and operate normally. A lag in any critical path activity will delay completion of the project, unless the project plan can be adjusted so that successor activity finishes more quickly than planned and cause abnormality of operation in the software project. The Critical Path Analysis is used as a decision making process for identifying the major components and the seriousness of their faults [8]. It identifies the maximum time for an operation to be completed.

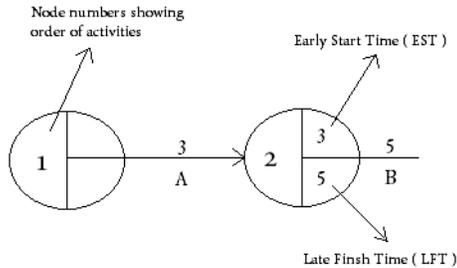

Fig. 2 Critical Activity Analysis

### 2.2 Graphical Representation

Graph constituting modules or dots have directed edges. The dependencies of modules are represented by forming an edge from module-m to module- n. This edge can be uni-directional or bi-directional depending on their relation. A directed edge from module- m to module- n implies that m is dependent on n. The key factor to look around are the critical dots which are essential to the software architect [2].

As explained above, a software project can be divided into modules and further sub-divided into activities. Equating activities to dots (.), our aim is to connect all the dots and form a coherent and meaningful relationship that describes the inter-dependencies of modules. We can achieve this by forming a dependency graph and dependency matrix using the concept of Graph theory [3] [5]. Dots (modules) are connected based on their dependencies on each other to form the graph. Critical activity is represented in a different manner than non-critical activity.

The vertices of the weighted graph can be considered as the part of the network and association between them denoted by the directed edge which forms the basis of dependency matrix [5]. Here assigned weight denotes the activity value between parts that is the node of the graph. In simple terms directed graph or digraph is a graph, or set of nodes connected by edges, where the edges have a direction associated with them [3]. In formal terms a graph is a pair vertices(V) and arcs (A)

- A set V, whose elements are called vertices or nodes

- A set A of ordered pairs of vertices, called arcs, directed edges, or arrows (and sometimes simply edges with the corresponding set named E instead of A)

Considering the following Scenario:

A sample case is taken to demonstrate graphical representation of activities.

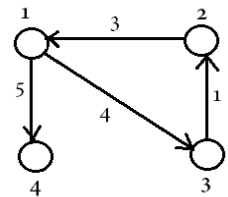

Fig. 3 Graphical Representation

So in this case we can find set of activities of software project shown in graphical manner dependent on each other with some weight assigned may be taken as time overhead or processing cost [2].

## 3. ANALYSIS

We are concerned about finding or say localizing the fault in a real time software project. Basic concept is used to find the bug which is done by using the concept of dependency matrix. Proposed approach is based on the critical activities of software project as modular approach is being followed in real time scenario. Large number of activities is split into small activities and later merged together to come up with desired output or operation.

### 3.1 Real Time Engineering

We have considered the scenario of working pattern of a Robot. Basic idea behind it is to find the module that causes malfunctioning of the robot. We have applied concepts of Graph Theory using dependency, incidence matrices to come to conclusion of identification of node affecting the working pattern [2].

Robot parts/modules are taken as nodes of the graph and the connection between them is considered as the edges between them with some weights denoting the synapse time of the individual parts. Here the values between nodes that are the active parts of the robot denotes the action synapse time. Now we are concerned about finding the critical nodes of activity. Critical nodes are the parts of the whole robot that has to be there for performing the particular activity.

On the basis of the critical nodes, we form the critical path that is minimized number of resources required in maximum allowed time to perform any given time.

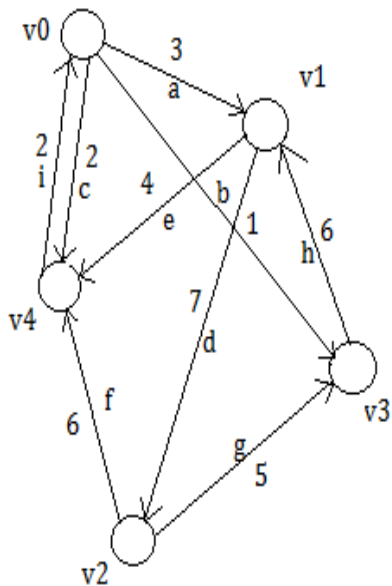

Fig. 4 Representation of Robot in graphical mode

So the robot arm we considered has the time limit of 15 sec as synapse response to perform basic amenities like lifting of object through the fore arms.

We can form the incidence matrix of the above graph obtained for the robot graph.

$$\begin{array}{c|ccccccccc} & a & b & c & d & e & f & g & h & i \\ \hline v_0 & 3 & 1 & 2 & 0 & 0 & 0 & 0 & 0 & 0 \\ v_1 & 0 & 0 & 0 & 7 & 4 & 0 & 0 & 0 & 0 \\ v_2 & 0 & 0 & 0 & 0 & 0 & 6 & 5 & 0 & 0 \\ v_3 & 0 & 6 & 0 & 0 & 0 & 0 & 0 & 0 & 0 \\ v_4 & 0 & 0 & 0 & 0 & 0 & 0 & 0 & 0 & 2 \end{array}$$

Fig. 5 Incidence Matrix of the Graph

Now we know that **v0, v1, v2 and v3** are the critical activities as the critical time path formed by the following nodes gives the critical time of 15 sec. The calculation of critical path can be depicted form the concept that there is no slack time for the performing the activities of the gives parts as nodes [9].

Now on the basis of the interaction of the parts of the robot, we can trace the dependency matrix of the graph using the precedence node of the corresponding nodes. This can be denoted in fig 6.

So in case of any project failure we can trace the activities getting affected as critical activities are the first to get affected.

$$\begin{array}{c|ccccc} & v_0 & v_1 & v_2 & v_3 & v_4 \\ \hline v_0 & 0 & 1 & 0 & 1 & 1 \\ v_1 & 0 & 0 & 1 & 0 & 1 \\ v_2 & 0 & 0 & 0 & 1 & 1 \\ v_3 & 0 & 1 & 0 & 0 & 0 \\ v_4 & 1 & 0 & 0 & 0 & 0 \end{array}$$

Fig. 6 Dependency Graph

Now by dependency principle we can trace the activities or nodes which are dependent on the critical nodes and hence it will be easy to monitor the performance of overall project using this concept of project handling. As we can reach to root node causing the failure to the software processing using dependency approach. It saves reasonable amount of time as general approach involves pipelining or random scan approach. We start looking from the critical nodes and then look for the dependant nodes so that time complexity can be minimized.

**3.2 Coding Instances**

We can come across multiple activities in big software project but when we dig inside, we come across bunch of actual codes which express the implementation of the logic and algorithm to actually operate the system. Looking carefully, we notice that apart from considering critical activity as such, we can also drill inside it and start off from our base.

Normally we come across software bug and code injection. It causes the software to react in unusual manner. Sometimes it can be easily traced by the programmers but most of the times it results unusual mess to handle. System that contains a large number of bugs, and/or bugs that fervently intrude with its functionality and operations, is said to be buggy. Reports enumerating bugs in a program are commonly known as bug reports, defect reports, fault reports, problem reports, and so forth [1]. Testers and debuggers have to go through all the code and multiple functions to trace the origin of the bug or say error to fix the system.

The process of debugging a software project may varies form interactive debugging or control flow manipulations. As we know that software systems have become generally more entangled, the various common debugging techniques have dilated with multiple methods to detect anomalies, estimate impact, and give remedy to a system for the given situation [1].

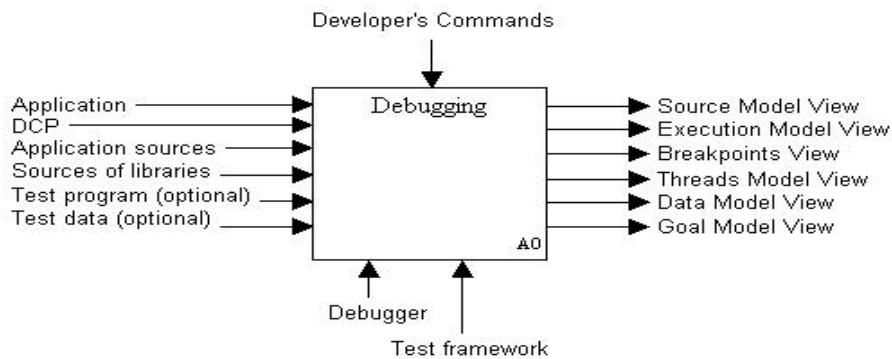

Fig. 7 Traditional Debugging Methodolgy

**3.3 Proposed Methodology**

We have critical activities when comes to real time systems for software project. Here we are going to talk about the critical section of the code which actually determines the outcome of the software system.
We usually have a lot of chunk of code in software development and it becomes quite complex for developer to keep track of all the functions he defined and also appropriate use of variables and often leads of code injection or some dead-recursion [4].

Now considering only the critical section, it saves the time thrice-fold or more. We can look for the dependency between the critical sections of the code. As mentioned earlier Software has multiple modules and hence here at crust level that is the code level we have multiple functions and constructors calling each other and get dependant on each other and other modules. So here basic idea lies behind directly looking for the critical section in the piece of code then tracking the dependant modules or other piece of code.

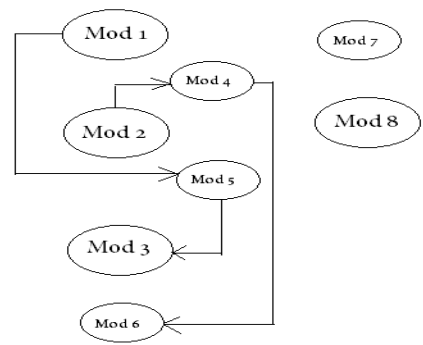

Fig. 8 Dependency between modules and sub-modules

So we can see the dependency of multiple modules in the above diagram. Some of them in critical section which actually plays the important role in the operation of logic and algorithm. So if we start tracing the bug in the code one should start with the critical section and then look for the dependant sections to minimize time overhead.

## 4. FUTURE WORKS

The work that we have done can be extended in many fields may it be hardware or software. Critical nodes are there in every kind of project and their essence can be felt while executing the same. Our work is novel and this can be extended to fields like biology and electronics. It can help to optimise the working environment and very easy to debug the system. People always get confused or messed up while figuring out the source of the defect. Using the mentioned approach, we can build an automated system that can help to easily resolve the anomalies in the software project.

## 5. RELATED WORKS

There has been lot of research in this field. We came across numerous approaches to find the source of bug in the system. Normally back-tracking or pipelining principle is followed to figure out the fault location but it consumes lot of time and resources plays important role in field of software development. Critical activities are always used by the project managers to determine to perform scheduling of the project as well as critical path analysis but here in this paper, critical activities are considered to find the bug or defect in the system as they play crucial part of the system.

## 6. CONCLUSION

The paper focuses on detecting the faults in a software project and finding the seriousness of the faults detected. We use back-tracking, divide and conquer or Branch and hound principle to track or localize the actual source of bug but it causes a lot of time. By considering only critical activities, time complexity can be managed as well as considering their dependant helps us to proceed in definite direction to reach the goal. This is done by observing the changes in the adjacency and dependency matrix formed from the weighted diagraph. Also, the critical path analysis helps in detecting the seriousness of these faults.